\documentclass[11pt]{article}
\usepackage[margin=0.75in]{geometry}
\usepackage[T1]{fontenc}
\usepackage[utf8]{inputenc}
\usepackage[english]{babel}
\usepackage{lmodern}
\usepackage{microtype}

\usepackage{mathtools,amssymb,amsthm}
\usepackage{mathrsfs,dsfont}
\usepackage{tikz-cd}
\usepackage{graphicx}
\usepackage{booktabs}
\usepackage{enumitem}
\usepackage{multicol}
\usepackage{authblk}

\usepackage{xcolor}

\usepackage{csquotes}

\usepackage[hidelinks]{hyperref}
\usepackage[english,nameinlink,noabbrev]{cleveref}

\theoremstyle{plain}
\newtheorem{theorem}{Theorem}[section]

\newtheorem{proposition}[theorem]{Proposition}
\newtheorem{corollary}[theorem]{Corollary}

\theoremstyle{definition}

\theoremstyle{remark}
\newtheorem{remark}[theorem]{Remark}

\crefname{theorem}{theorem}{theorems}
\Crefname{theorem}{Theorem}{Theorems}
\crefname{lemma}{lemma}{lemmas}
\Crefname{lemma}{Lemma}{Lemmas}
\crefname{proposition}{proposition}{propositions}
\Crefname{proposition}{Proposition}{Propositions}
\crefname{corollary}{corollary}{corollaries}
\Crefname{corollary}{Corollary}{Corollaries}
\crefname{conjecture}{conjecture}{conjectures}
\Crefname{conjecture}{Conjecture}{Conjectures}
\crefname{definition}{definition}{definitions}
\Crefname{definition}{Definition}{Definitions}
\crefname{example}{example}{examples}
\Crefname{example}{Example}{Examples}
\crefname{notation}{notation}{notations}
\Crefname{notation}{Notation}{Notations}
\crefname{remark}{remark}{remarks}
\Crefname{remark}{Remark}{Remarks}

\newcommand{\ii}{\mathrm{i}}
\newcommand{\PP}{\mathcal{P}}
\newcommand{\RR}{\mathcal{R}}
\newcommand{\hS}{\hat{S}}

\title{Bi-Hamiltonian structure and conserved quantities of the Yajima--Oikawa--Newell system}
\author{Marcos Caso-Huerta}
\affil{Department of Mathematics, University of Oviedo, C/ Leopoldo Calvo-Sotelo 18, 33007 Oviedo, Spain}

\date{}

\begin{document}
\maketitle

\begin{abstract}
We study the Hamiltonian and bi-Hamiltonian structure of the Yajima--Oikawa--Newell (YON) system, an integrable model of long wave--short wave interaction depending on two real parameters $\alpha$ and $\beta$, which reduces to the Yajima--Oikawa system for $\alpha=0$ and to the Newell system for $\beta=0$. For $\alpha\neq0$, a constant shift of the long wave maps the YON system into the Newell system, for which two nonlocal Hamiltonian operators are known. We show that, for $\alpha\neq0$, the system admits no local Hamiltonian operator homogeneous in its natural grading and polynomial in $\beta$: at each admissible weight, the weight and $U(1)$-charge constraints leave a finite-dimensional family of candidates, none of which is a Poisson operator. A nonlocal gauge transformation built on the potential of the long wave makes the first of the known operators local for a single value of its parameter: the transformed system is Hamiltonian, for every $\alpha\neq0$ and every $\beta$, with respect to a local, non-degenerate Poisson operator, which is the Lie--Poisson structure of the $\mathfrak{sl}(2)$ current algebra. From a recursion operator advancing the hierarchy by two steps, given in explicit weakly nonlocal form, we construct a second, weakly nonlocal Hamiltonian structure, whose nonlocal tail is generated by the three autonomous Lie point symmetries of the system, and prove that both structures are compatible. Since the recursion operator admits no square root and the first structure has no Casimirs, the Lenard--Magri chain splits into two families, which have to be seeded by the two simplest conserved quantities. We show that every flow of both families is local and that all the corresponding Hamiltonians are in involution. As by-products, we relate the Lie point symmetries of the system to its conserved quantities via Noether's theorem, and obtain the conserved densities in closed form by asymptotic diagonalisation of the Lax pair.
\end{abstract}

\section{Introduction}
\label{sec:introduction}
The resonant interaction between a long wave and a short wave, which takes place whenever the group velocity of the short wave matches the phase velocity of the long wave \cite{Benney1977}, is one of the classical mechanisms of nonlinear wave theory, and two of its best-known integrable models are the Yajima--Oikawa (YO) system \cite{Yajima1976,Djordjevic1977,Ma1978} and the Newell system \cite{Benney1977,Newell1978}. In \cite{CasoHuerta2021}, both models were unified and generalised into the Yajima--Oikawa--Newell (YON) system
\begin{subequations}
\label{YON}
    \begin{align}
        & \ii S_t + S_{xx} + (\ii \alpha L_x + \alpha^2 L^2 - \beta L - 2\alpha |S|^2)S=0, \label{YONS}\\
        & L_t = 2(|S|^2)_x, \label{YONL}
    \end{align}
\end{subequations}
where $S$ is a complex variable representing the short-wave envelope, $L$ is the real amplitude of the long wave, and $\alpha$ and $\beta$ are two real parameters.

The YON system is integrable, in the sense of possessing a Lax pair, for every choice of $\alpha$ and $\beta$ \cite{CasoHuerta2021}, and it contains the two classical models as particular cases. Namely, taking $\alpha=0$ and $\beta=1$ in Eqs.~\eqref{YON} we recover the YO system
\begin{subequations}
    \label{YajimaOikawa}
    \begin{align}
        & \ii S_t + S_{xx} - LS = 0,\\
        & L_t = 2(|S|^2)_x,
    \end{align}
\end{subequations}
whereas taking $\alpha=\sigma$, $\beta=0$ and $L\mapsto \sigma L$, with $\sigma=\pm 1$, we recover the Newell system
\begin{subequations}
    \label{Newell}
    \begin{align}
        & \ii S_t + S_{xx} + (\ii L_x + L^2 - 2\sigma|S|^2)S=0,\\
        & L_t = 2\sigma(|S|^2)_x.
    \end{align}
\end{subequations}
Let us remark that the choices $\beta=1$ and $\alpha=\sigma$ entail no loss of generality: simple scalings of the fields and of the independent variables allow one to normalise any nonzero value of $\alpha$ to $\pm1$ and any nonzero value of $\beta$ to $1$ (see Section~\ref{SecNoGo}). In fact, for $\alpha\neq0$ the parameter $\beta$ can be removed altogether: a constant shift of $L$, together with a time-dependent phase of $S$, maps the YON system into the same system with $\beta=0$, that is, into the Newell system, at the price of replacing the vanishing boundary condition of $L$ by a constant background. We will nevertheless keep both parameters arbitrary throughout, as we consider fields that vanish at infinity, and as this makes both the grading of the system and its reductions to YO and Newell more transparent.

Since its introduction, the YON system and its solutions have attracted steady interest. The linear stability of its plane wave solutions was analysed in \cite{CasoHuerta2021}, and its solitons, breathers and rogue waves were obtained by an ansatz approach in \cite{CasoHuerta2022} and within the Hirota bilinear framework in \cite{Kirane2022,CasoHuerta2025}, and further studied in \cite{Bai2025,Ankiewicz2026}. In \cite{CasoHuerta2026}, the YON system is derived from physical principles, by applying the multiscale method to a modified KdV equation for shallow water waves, and in \cite{Myrzakulova2023} it appears in connection with the Tolkynay equations.

Some conservation laws of Eqs.~\eqref{YON} were obtained in \cite{CasoHuerta2022} by means of the multiplier method of Anco and Bluman \cite{Anco1997,Anco2002}, as no variational formulation of the YON system was known. For its two reductions, on the other hand, a Hamiltonian theory is available. The YO system \eqref{YajimaOikawa} is well known to be bi-Hamiltonian, being a reduction of the constrained KP hierarchy \cite{Oevel1993,Li2015}, and Liu \cite{Liu1994} showed that the Newell system \eqref{Newell} is a modification of it, the two systems being related by a Miura-type map, and gave two Hamiltonian operators for the Newell hierarchy. Both of them are nonlocal in the variables $(S,S^*,L)$: the first one is weakly nonlocal, with a tail built on the phase symmetry, whereas the second one involves inverses of differential operators. To the best of our knowledge, these results have not been related to the YON system, and it is not known whether the nonlocality of these operators can be avoided. The purpose of this paper is to address both questions.

As we will see, the nonlocality is unavoidable in the original variables, but it is not intrinsic. For $\alpha\neq0$, the shift of $L$ mentioned above reduces the YON system to the Newell system, so that the operators of \cite{Liu1994} apply to it. We first show that the natural local candidates, namely operators that are homogeneous in the grading of the system and compatible with its $U(1)$ invariance, all fail to be Poisson as soon as $\alpha\neq0$. A phase gauge on the potential of the long wave, very much in the spirit of the gauge transformations relating the members of the derivative NLS family \cite{Kaup1978,Chen1979,Gerdjikov1983,Wadati1983,Kundu1984}, then turns the first operator of \cite{Liu1994} into a local one, and it does so for a single value of the gauge parameter. In that gauge the operator is linear in the fields, and it is in fact the Lie--Poisson structure of the $\mathfrak{sl}(2)$ current algebra, which makes its Jacobi identity manifest. Taking this local, non-degenerate operator as the first structure, we construct a recursion operator for the hierarchy in explicit, weakly nonlocal form and, from it, a second Hamiltonian structure, also weakly nonlocal, which lies one step above the pair of \cite{Liu1994} and whose nonlocal tail is generated by the three autonomous Lie point symmetries of the system. The resulting picture has some features that are worth pointing out from the start. The recursion operator advances the hierarchy by two steps rather than one, as in the YO case, and admits no square root, so that the hierarchy splits into two families; and, since the first structure has no Casimirs, the Lenard--Magri chain cannot be started, as usual, from a Casimir, but has to be seeded by the two simplest conserved quantities.

The paper is organised as follows. In Section~\ref{SecNoGo}, we recall the Lax pair of the YON system, introduce its grading and $U(1)$ charges, and use them to prove that, for $\alpha\neq0$, the system admits no local Hamiltonian operator homogeneous in that grading and polynomial in $\beta$. In Section~\ref{SecNonlocal}, we show that a nonlocal gauge transformation on the potential of the long wave makes the system Hamiltonian with respect to a local, non-degenerate operator, which we identify both with the first operator of \cite{Liu1994} and with the Lie--Poisson structure of the $\mathfrak{sl}(2)$ current algebra, construct the conserved densities in closed form by asymptotic diagonalisation of the Lax pair, and take advantage of the Hamiltonian formulation to relate the Lie point symmetries of the system obtained in \cite{CasoHuertaThesis} to its conserved quantities via Noether's theorem. In Section~\ref{SecSecond}, we construct a recursion operator acting on the symmetries of the system and, from it, a second Hamiltonian structure, which is weakly nonlocal in the sense of Ferapontov, Maltsev and Novikov \cite{Ferapontov1991,Maltsev2001,Casati2020}, and prove that both structures are compatible. In fact, we show that the nonlocal tail is generated not just by symmetries of the system, but by its three autonomous Lie point symmetries. In Section~\ref{SecLenard}, we study the Lenard--Magri chain of the pair, show that every flow of both families is local and that the corresponding Hamiltonians are in involution, and discuss the reductions $\beta=0$ and $\alpha=0$. Finally, Section~\ref{SecConclusions} collects our conclusions.

\section{No-go theorem for local Hamiltonian structures}\label{SecNoGo}
\subsection{Lax pair, symmetries and grading}
Before looking for a Hamiltonian structure, let us collect the properties of the YON system that will constrain the search, namely its Lax pair, its scaling symmetries and the grading they induce, and its $U(1)$ invariance. The YON system \eqref{YON} arises as the compatibility condition $X_t-T_x+[X,T]=0$ of the Lax pair \cite{CasoHuerta2021}
\begin{subequations}\label{LaxPair}
\begin{equation}
    X(\lambda)=\ii\lambda\Sigma+Q,\qquad T(\lambda)=(\ii\lambda)^2T_2 + (\ii\lambda) T_1 + T_0, \qquad \Sigma=\text{diag}(1,0,-1),
\end{equation}
with
\begin{equation}
\label{YONQ}
Q=\left (\begin{array}{ccc} 0 & S & \ii L \\ \alpha  S^\ast & 0 &   S^\ast  \\ \ii\alpha^2L-\ii\beta & \alpha S & 0 \end{array} \right )\,,
\end{equation}
\begin{equation}
\label{YONT12}
T_2=\frac{\ii}{3}\left (\begin{array}{ccc} 1 & 0 & 0 \\ 0 & -2 & 0  \\0 & 0 & 1 \end{array} \right ),
\quad T_1= \left (\begin{array}{ccc} 0 & \ii S & 0 \\ \ii \alpha S^\ast & 0 & -\ii S^\ast  \\ 0 & -\ii\alpha S & 0 \end{array} \right ),
\end{equation}
\begin{equation}
\label{YONT0}
T_0= \left (\begin{array}{ccc} -\ii\alpha |S|^2 & -\alpha LS +\ii S_x & \ii|S|^2 \\ - \alpha^2 LS^\ast +\beta  S^\ast-\ii \alpha S^\ast_x  & 2\ii \alpha |S|^2 &  -\alpha LS^\ast -\ii S^\ast_x \\ \ii \alpha^2 |S|^2& -\alpha^2 LS +\beta S+\ii\alpha S_x & -\ii\alpha |S|^2 \end{array} \right ).
\end{equation}
\end{subequations}

Eqs. \eqref{YON} are invariant under the scaling
\begin{equation}
\label{scaling1}
    (S,L,x,t,\alpha,\beta)\mapsto (\varepsilon S, \varepsilon L, \varepsilon^{-1}x, \varepsilon^{-2}t,\alpha,\varepsilon\beta),
\end{equation}
which induces the grading
\begin{equation}
\label{grading}
    [S]=[S^*]=[L]=[\partial_x]=[\beta]=1,\qquad [\alpha]=0,\qquad [\partial_t]=2,
\end{equation}
in which the flow \eqref{YON} is homogeneous of weight 3. Let us note in passing that Eqs. \eqref{YON} are also invariant under the transformation
\begin{equation}
\label{scaling2}
    (S,L,\alpha,\beta)\mapsto (aS,a^2L,a^{-2}\alpha,a^{-2}\beta),\qquad a\in\mathbb{R}\setminus\{0\},
\end{equation}
which leaves $x$ and $t$ unchanged. Together, the two scalings allow one to normalise any nonzero value of $\alpha$ to $\pm1$ and any nonzero value of $\beta$ to $1$, as anticipated in the Introduction. Moreover, for $\alpha\neq0$ the parameter $\beta$ can be removed altogether: if $(S,L)$ solves Eqs. \eqref{YON}, then
\begin{equation}
\label{shift}
    \tilde S = \exp\left(\frac{\ii\beta^2 t}{4\alpha^2}\right)S,\qquad \tilde L = L - \frac{\beta}{2\alpha^2}
\end{equation}
solve the same equations with $\beta=0$, that is, the Newell system. In other words, for $\alpha\neq0$ the parameter $\beta$ plays the role of a constant background of the long wave in the Newell system. We will not make use of any of these freedoms: the transformation \eqref{shift} does not preserve the vanishing of $L$ at infinity, and the grading \eqref{grading} relies precisely on $\beta$ being a parameter of weight one.

Let us now see how the grading constrains a Hamiltonian formulation. We will restrict ourselves to operators and densities that are homogeneous in the grading \eqref{grading} and polynomial in $\beta$, in the fields and in their $x$-derivatives, with coefficients that may depend on $\alpha$. If $\PP$ is such a Hamiltonian operator and $H=\int h\,dx$ is the corresponding Hamiltonian, the relation $u_t=\PP\,\delta H$, with $u=(S,S^*,L)$, implies that
\begin{equation}
\label{weightrule}
    3=\text{weight}(\PP_{ab}) + [h] - 1
\end{equation}
for each entry $\PP_{ab}$ of the operator, since all three fields have weight one and the variational derivative lowers the weight by one. Hence, $\PP$ must be homogeneous of weight $p=4-[h]$. Since the entries of $\PP$ have non-negative weight, $p\geq0$, while $p\geq4$ would require $[h]\leq0$, that is, a constant density, for which $\delta H=0$. Only the weights $p=0,1,2,3$ remain, and for each of them the space of candidates is finite-dimensional.

We can further restrict the class of candidates by means of the $U(1)$ symmetry $S\mapsto e^{\ii \phi}S$, $\phi\in\mathbb{R}$. Assigning charges $+1$, $-1$ and $0$ to $S$, $S^*$ and $L$, respectively, the entry $\PP_{ab}$ must carry charge $q_a + q_b$, since $\PP$ maps covectors, whose components $\delta H/\delta u_b$ carry charge $-q_b$, to vectors, whose components carry charge $q_a$. Combining this condition with the skew-adjointness of the operator and with the reality condition (that complex conjugation exchanges the roles of $S$ and $S^*$ and fixes $L$), the problem at each admissible weight reduces to a linear problem for finitely many coefficients.

It remains, therefore, to analyse the weights 0, 1, 2 and 3. Since the Hamiltonian density, of weight $[h]=4-p$, has to be conserved, it is convenient to compute first all the conserved densities of weights one to three, by solving $\delta(\partial_t\rho)=0$ on the graded monomial basis of each weight. Modulo total derivatives, this results in
\begin{equation}
\label{densities}
    \rho_0 = L,\quad \rho_1 = \frac{\alpha}{2}L^2 - |S|^2,\quad \rho_2 = 2\alpha^2 L |S|^2-\beta|S|^2 + 2\alpha\Im(S^*S_x),
\end{equation}
which, apart from the products of the lower ones by powers of $\beta$, are the only conserved densities of their weight, and which match the ones derived in \cite{CasoHuerta2022}. Here and in what follows, for a density $\rho$ we write $\delta\rho$ for the variational derivative of $\int\rho\,dx$. A single new density appears at weight four as well, and we will come back to it in Section \ref{SecNonlocal}.

\subsection{Search for local Hamiltonian operators}
With the candidates reduced to a finite-dimensional family at each weight, let us now determine which of them, if any, are Poisson operators. To this end, we will check the Jacobi identity via the Schouten criterion \cite{Olver1993,Dorfman1993}, that is, a skew-adjoint operator $\PP$ is Hamiltonian if and only if, given arbitrary covectors $\theta^1$, $\theta^2$, $\theta^3$, the Schouten bracket \cite{Schouten1940,Schouten1954,Nijenhuis1955}
\begin{equation}
\label{Schouten}
    [\PP,\PP]=\sum_{\text{cyc}} \langle \theta^1, \PP'[\PP\theta^2]\theta^3 \rangle,
\end{equation}
is a total $x$-derivative, where the sum is extended over all cyclic permutations of the indices, $\langle\theta,v\rangle=\sum_a\theta_a v_a$ is the bilinear covector-vector pairing, with no complex conjugation, $S$ and $S^*$ being treated as independent fields, and $\PP'[w]$ is the Fr\'echet derivative of the coefficients of $\PP$ along the direction of $w$. Whether an expression is a total $x$-derivative is decided, in turn, by the Euler operator: it is so if and only if its variational derivatives with respect to all the fields and all the components of the covectors vanish identically.

Weight zero is excluded by the $U(1)$-charge and skew-adjointness conditions alone. Indeed, the entries are then constants, and the charge rule allows only for $\PP_{12}$, $\PP_{21}$ and $\PP_{33}$; but a constant $\PP_{33}$ is self-adjoint, so skew-adjointness forces it to vanish. Hence, the last row of $\PP$ vanishes identically, and the operator cannot reproduce the equation for $L_t$.

\begin{theorem}[Weight one]
\label{TheoremWeightOne}
    Let $\PP^{(1)}$ be a skew-adjoint operator, homogeneous of weight 1 under the grading \eqref{grading}, satisfying the reality and $U(1)$-charge conditions, which generates the flow \eqref{YON} with a Hamiltonian whose density is conserved of weight three. Up to multiplication by a constant, $\PP^{(1)}$ is forced to take the form
    \begin{equation}
        \PP^{(1)}=\begin{pmatrix}
            0 & -\alpha\partial_x + \ii \alpha^2 L - \dfrac{\ii \beta}{2} & -2\ii \alpha S \\
            -\alpha\partial_x - \ii \alpha^2 L + \dfrac{\ii \beta}{2} & 0 & 2\ii \alpha S^* \\
            2\ii \alpha S & -2\ii \alpha S^* & 4 \partial_x
        \end{pmatrix}.
    \end{equation}
    The corresponding Hamiltonian density is
    \begin{equation}
        h^{(1)} = \frac{1}{\alpha} \Im(S^* S_x) + L|S|^2 - \dfrac{\beta}{2\alpha^2} |S|^2 + \dfrac{\beta^2}{8\alpha^3} L,
    \end{equation}
    and its Schouten bracket \eqref{Schouten} reduces, modulo total derivatives, to a single trivector
    \begin{equation}
    [\PP^{(1)},\PP^{(1)}]\equiv -6\ii \alpha^2\sum_\text{cyc} (\theta_3^1)_x(\theta_1^2\theta_2^3-\theta_2^2\theta_1^3),
    \end{equation}
    where the right-hand side is not a total derivative.

    Hence, $\PP^{(1)}$ is a Poisson operator if and only if $\alpha=0$. Note, however, that at $\alpha=0$ it reduces to the constant-coefficient operator $\text{diag-block}\{\frac{\beta}{2}\sigma_2,4\partial_x\}$, with $\sigma_2$ the second Pauli matrix, which, although Poisson, generates the Yajima--Oikawa flow from no Hamiltonian at all, as the density $h^{(1)}$ is singular there. The Hamiltonian structure of the Yajima--Oikawa system is a different one, which we will recover in Section \ref{SecLenard}.
\end{theorem}

\begin{theorem}[Weight two]
\label{TheoremWeightTwo}
    The family of skew-adjoint operators, homogeneous of weight 2 under the grading \eqref{grading}, satisfying the reality and $U(1)$-charge conditions, which generate the flow \eqref{YON} with Hamiltonian $\int\rho_1\,dx$, takes the form, up to multiplication by a constant,
    \begin{subequations}
        \begin{equation}
            \PP^{(2)} = \begin{pmatrix}
                0 & \PP_{12}^{(2)} & \PP_{13}^{(2)} \\
                -{\PP_{12}^{(2)}}^\dagger & 0 & \overline{\PP_{13}^{(2)}} \\
                -{\PP_{13}^{(2)}}^\dagger & -{\overline{\PP_{13}^{(2)}}}^\dagger & 0
            \end{pmatrix},
        \end{equation}
        where
        \begin{equation}
            \begin{split}
                & \PP_{12}^{(2)} = - \ii \partial_x^2 + d\alpha L \partial_x + \ii (1+\alpha b)\beta L + \ii \alpha (a-\alpha)L^2 + 2\ii \alpha |S|^2 + \alpha \left(\ii c + \frac{d}{2}\right) L_x, \\
                & \PP_{13}^{(2)} = \left( -1 + \ii c + \frac{d}{2}\right) S \partial_x + d S_x + \ii (aL+b\beta) S,
            \end{split}
        \end{equation}
    \end{subequations}
    with $a,b,c,d\in\mathbb{R}$ arbitrary parameters. The associated Jacobi system, made of 287 distinct polynomial equations in $(a,b,c,d)$, is inconsistent for every value of $\alpha$ and $\beta$; hence, no member of the family is a Poisson operator.
\end{theorem}

\begin{remark}
    Admitting a Hamiltonian density that combines the two weight-two conserved densities $\rho_1$ and $\beta\rho_0$ extends the class of candidates to a total of five parameters (with 310 distinct equations in the Jacobi system), but it does not change the conclusion of Theorem \ref{TheoremWeightTwo}.
\end{remark}

The last remaining weight is $p=3$, which implies $[h]=1$ and hence the Hamiltonian $H=\int L \,dx$, $\rho_0$ being the only weight-one density other than the constant $\beta$. For this choice, the deciding test is not the Jacobi identity but a simpler necessary condition. Since a Hamiltonian vector field must preserve its own Poisson structure, any Poisson operator $\PP$ that generates Eqs. \eqref{YON} must satisfy
\begin{equation}
\label{LieW3}
    \mathcal{L}_K \PP = \PP'[K]-K'\PP - \PP(K')^\dagger = 0,
\end{equation}
where $\mathcal{L}$ denotes the Lie derivative, $K$ is the right-hand side of Eqs. \eqref{YON}, $K'$ is its Fr\'echet derivative and $\PP'[K]$ is the derivative of the coefficients of $\PP$ along $K$. Relation \eqref{LieW3} is a linear identity, allowing us to produce the following result.
\begin{theorem}[Weight three]
\label{TheoremWeightThree}
    The family of skew-adjoint operators, homogeneous of weight 3 under the grading \eqref{grading}, satisfying the reality and $U(1)$-charge conditions and the identity $\PP\delta\int L\,dx=K$ is a 31-parameter affine family. Introducing the linear condition \eqref{LieW3} on the general 31-parameter operator results in a linear system of rank 31 and augmented rank 32, meaning that the system is incompatible. The incompatibility holds for every value of $\alpha$ and $\beta$. Hence, no member of the family is a Poisson operator.
\end{theorem}

Since we have obtained a negative answer at every admissible weight, we can condense the results in the following:
\begin{corollary}
\label{CorNoGo}
    For $\alpha\neq0$, the YON system \eqref{YON} admits no local Hamiltonian structure in the variables $(S,S^*,L)$ whose operator and Hamiltonian density are homogeneous in the grading \eqref{grading} and polynomial in $\beta$.
\end{corollary}
\begin{remark}
\label{RemYOclass}
    The case $\alpha=0$ is not an exception within this class. The canonical Hamiltonian operator of the YO system, which we will recover in Section \ref{SecLenard}, is homogeneous of weight one, but its Hamiltonian contains the term $|S_x|^2/\beta$, and is therefore not polynomial in $\beta$. This reflects the fact that the natural grading of the YO system is a different one, $[S]=3/2$, $[L]=2$ and $[\beta]=0$, which for the YON system results from combining the scalings \eqref{scaling1} and \eqref{scaling2} and assigns weight $-1$ to $\alpha$. For $\alpha\neq0$, on the other hand, admitting such terms does not change the conclusion at weight one: with densities containing negative powers of $\beta$ down to $\beta^{-2}$, the operator of Theorem \ref{TheoremWeightOne} remains the only candidate.
\end{remark}

Having exhausted the homogeneous local candidates, we are thus led to look for a Hamiltonian structure beyond the local class in the original variables. This is consistent with the results of \cite{Liu1994} for the Newell system, where both Hamiltonian operators are nonlocal in these variables. As we will see in the next section, however, the nonlocality of the first of them can be removed by a change of variables.


\section{Nonlocal Hamiltonian structure}\label{SecNonlocal}
Section \ref{SecNoGo} leaves us with a negative answer in the original variables. Let us now show that the obstruction found there is tied to the choice of variables, and that it can be removed by a nonlocal change of them. A natural candidate is suggested by the Newell system itself: in \cite{Newell1978}, Newell showed that a phase gauge whose exponent is the potential of the long wave simplifies system \eqref{Newell} considerably \cite{CasoHuertaThesis}. Let $W$ satisfy $W_x=L$ and $W_t=2|S|^2$, as allowed by Eq. \eqref{YONL}, and consider a gauge transformation of the form $S=\exp(-\ii \gamma W)\,\hat{S}$, with $\gamma$ a real parameter. Introducing the transformation in Eq. \eqref{YONS} yields
\begin{equation}
\label{GaugeFlow}
    \ii \hS_t + \hS_{xx} - 2\ii\gamma L\hS_x + \ii (\alpha-\gamma)L_x \hS + (\alpha^2-\gamma^2)L^2\hS - \beta L\hS + 2(\gamma-\alpha)|\hS|^2 \hS = 0.
\end{equation}

Repeating the procedure of Theorem \ref{TheoremWeightOne} with the flow \eqref{GaugeFlow} leads to a Schouten bracket that reduces, modulo total derivatives, to a trivector
\begin{equation}
    [\PP,\PP] \equiv -2\ii (\alpha-\gamma)(3\alpha+\gamma)\sum_\text{cyc} (\theta_3^1)_x(\theta_1^2\theta_2^3-\theta_2^2\theta_1^3).
\end{equation}
The root $\gamma=\alpha$ here is spurious: there, the whole short-wave part of the operator carries the factor $\alpha-\gamma$ and vanishes, so that $\PP$ reduces to $\text{diag}(0,0,4\partial_x)$ and fails to generate the flow. This is precisely the gauge of Newell, in which Eq. \eqref{GaugeFlow} becomes linear in $\hS$, namely $\ii\hS_t+\hS_{xx}-2\ii\alpha L\hS_x-\beta L\hS=0$; the gauge that linearises the short-wave equation is therefore not the one that makes the system Hamiltonian in local form. This leaves $\gamma=-3\alpha$ as the only genuine root within the weight-one class. Let us remark that the restriction on $\gamma$ is a restriction on the gauge transformation and not on the YON system: the transformed system \eqref{GaugeFlow} with $\gamma=-3\alpha$ can be endowed with a local Hamiltonian structure, and therefore the YON system \eqref{YON} admits a nonlocal Hamiltonian structure for every $\alpha\neq0$ and $\beta$ via the gauge transformation $S=\exp(3\ii\alpha W)\,\hS$.

\subsection{Explicit Hamiltonian structure and hierarchy}
Introducing the choice $\gamma=-3\alpha$ in Eq. \eqref{GaugeFlow} leads to the system
\begin{subequations} \label{LocalYON}
    \begin{align}
        & \ii \hS_t + \hS_{xx} + 6\ii \alpha L \hS_x + 4\ii \alpha L_x \hS - 8\alpha^2 L^2 \hS - \beta L\hS - 8\alpha|\hS|^2\hS = 0,\\
        & L_t = 2(|\hS|^2)_x.
    \end{align}
\end{subequations}
We can introduce its Hamiltonian structure through the following result.
\begin{theorem}
\label{ThmPstar}
    The skew-adjoint operator
    \begin{equation}
    \label{P*}
        \PP_* = \begin{pmatrix}
            0 & -\alpha\partial_x - \ii \mu & -2\ii\alpha \hS \\
            -\alpha\partial_x + \ii \mu & 0 & 2\ii\alpha \hS^* \\
            2\ii\alpha \hS & -2\ii\alpha \hS^* & \partial_x
        \end{pmatrix},
    \end{equation}
    where $\mu=2\alpha^2 L + \dfrac{\beta}{2}$, is a Hamiltonian operator, and system \eqref{LocalYON} is its Hamiltonian flow $u_t=\PP_* \,\delta H_*$, generated by
    \begin{equation}\label{H*}
        H_* = \int \left[ 4 L |\hS|^2 - \dfrac{\beta}{2\alpha^2} |\hS|^2 + \dfrac{1}{\alpha}\Im(\hS^*\hS_x) + \dfrac{\beta^2 L}{8\alpha^3} \right]\,dx,
    \end{equation}
    for every $\alpha\neq 0$ and $\beta$.
\end{theorem}

The Jacobi identity of $\PP_*$ can be understood in purely algebraic terms, since the operator is of first order and affine in the fields. Indeed, writing $\tilde u=(\hS,\hS^*,L+\beta/4\alpha^2)$, we have
\begin{equation}
\label{CurrentAlgebra}
    (\PP_*)_{ab} = g_{ab}\,\partial_x + \sum_{c=1}^3 f_{ab}^{\;c}\,\tilde u_c,\qquad
    g=\begin{pmatrix} 0 & -\alpha & 0 \\ -\alpha & 0 & 0 \\ 0 & 0 & 1 \end{pmatrix},
\end{equation}
where the only nonzero constants $f_{ab}^{\;c}=-f_{ba}^{\;c}$ are $f_{12}^{\;3}=-2\ii\alpha^2$, $f_{13}^{\;1}=-2\ii\alpha$ and $f_{23}^{\;2}=2\ii\alpha$. These are the structure constants of the Lie algebra $\mathfrak{sl}(2,\mathbb{C})$, and $g$ is an invariant form on it, namely $-1/8\alpha^2$ times its Killing form. Hence, $\PP_*$ is the Lie--Poisson structure of the centrally extended $\mathfrak{sl}(2)$ current algebra \cite{Drinfeld1985}, and its Jacobi identity reduces to that of the Lie algebra together with the invariance of $g$. Three consequences are worth noting. First, the parameter $\beta$ enters $\PP_*$ only through the shift of $L$ in $\tilde u$, in agreement with Eq. \eqref{shift}. Second, in the real coordinates $(\operatorname{Re}\hS,\operatorname{Im}\hS,L)$ the form $g$ becomes $\text{diag}(-\alpha/2,-\alpha/2,1)$, so that the real form of the algebra is the compact one, $\mathfrak{su}(2)$, for $\alpha<0$, and $\mathfrak{sl}(2,\mathbb{R})$ for $\alpha>0$: the two signs of $\alpha$, that is, the two signs $\sigma$ of the Newell system \eqref{Newell}, correspond to different real forms. Third, the quadratic functional $\frac12\int\langle\tilde u,g^{-1}\tilde u\rangle\,dx$ generates the translation, as in the Sugawara construction.

Transforming the local Poisson operator \eqref{P*} back to the original variables $(S,S^*,L)$ by means of the gauge transformation, we obtain a nonlocal Poisson structure for the YON system \eqref{YON}, with a weakly nonlocal tail built on the phase symmetry. For $\beta=0$, this is the first Hamiltonian operator given in \cite{Liu1994} for the Newell hierarchy: the modified system of \cite[Eq. (2.22)]{Liu1994}, in the variables $(\hat u,\hat q,\hat r)$ and time $t_2$ used there, is the Newell system under the identification $\hat u=\ii\alpha L$, $\hat q=aS$, $\hat r=2\alpha S^*/a$ and $t_2=2\ii t$, with $a$ an arbitrary nonzero constant, and, on covectors that decay at infinity, the operator of \cite[Eq. (2.17)]{Liu1994} coincides with $-1/2\alpha^2$ times the operator $\PP_*$ carried back to the original variables. The gauge $\gamma=-3\alpha$ thus provides the variables in which that operator becomes local, and Eq. \eqref{CurrentAlgebra} identifies it as a current algebra, whereas in \cite{Liu1994} it is reduced to a constant-coefficient operator by two further, non-invertible, Miura-type maps. Together with Eq. \eqref{shift}, this extends the first Hamiltonian structure of the Newell hierarchy to the YON system for every $\beta$.

The fact that a Hamiltonian structure can only be achieved after a nonlocal gauge transformation is not exclusive to YON. In fact, the mechanism here is similar to the one at work in the derivative NLS family. The Kaup--Newell, Chen--Lee--Liu and Gerdjikov--Ivanov equations \cite{Kaup1978,Chen1979,Gerdjikov1983} are related by phase gauges whose exponent is the potential of a conserved density \cite{Wadati1983,Kundu1984,Hisakado1994}, in the same way as $W$ is the potential of the conserved density $\rho_0=L$ in the YON case; there, the gauge parameter changes the Hamiltonian operator and the Hamiltonian together, and it can be chosen according to the member of the family one wants to display. The gauge-equivalent forms of the Newell system, also known as the derivative Yajima--Oikawa system, have themselves been classified into three types in analogy with the derivative NLS family \cite{Chen2018}. The key difference in the YON case is that $\gamma$ is not a choice: the Jacobi obstruction fixes $\gamma=-3\alpha$ as the only value for which a local Poisson operator exists.

Before studying the hierarchy generated by $\PP_*$, let us fix some notation for its lowest members. We denote by $K_{-1}=(\ii \hS,-\ii\hS^*,0)$ the phase symmetry, by $K_0=(\hS_x,\hS_x^*,L_x)$ the translation and by $K_1$ the flow \eqref{LocalYON}. In general, $\PP_*$ maps a Hamiltonian of weight $p$ to a flow of weight $p$, and the Hamiltonians of these three flows, $\PP_*\delta H_n=K_n$, are $H_{-1}=-\int L/2\alpha\,dx$, $H_0=\int(\rho_1/\alpha+\beta\rho_0/4\alpha^2)\,dx$ and $H_1=H_*$.

Let us now move on to study the degeneracy of the Poisson operator $\PP_*$.

\begin{theorem}[Non-degeneracy]
\label{NondegP*}
    Provided $\alpha\neq 0$, the operator $\PP_*$ is injective on local covectors, i. e., if the components of $g$ are differential polynomials in $(\hS,\hS^*,L)$ and $\PP_* g=0$, then $g=0$. In particular, $\PP_*$ has no nontrivial Casimir at any weight.
\end{theorem}
\begin{proof}
    Using Eq. \eqref{P*}, the equation $\PP_* g =0$ reads
    \begin{equation}
    \label{ProofP*}
        \alpha g_{2,x}=-\ii \mu g_2 - 2\ii\alpha \hS g_3,\qquad \alpha g_{1,x} = \ii \mu g_1 + 2\ii\alpha \hS^* g_3,\qquad g_{3,x} = -2\ii \alpha(\hS g_1-\hS^* g_2).
    \end{equation}
    Note that each of the three equations features exactly one $x$-derivative on the left-hand side and none on the right-hand side.

    Let $n_j$ be the order of $g_j$ as a differential polynomial, where $j=1,2,3$, and $N=\max_j n_j$, attained at $j=j_0$.

    Let us consider the equation corresponding to $g_{j_0,x}$. On the left-hand side, $g_{j_0,x}$ is a differential polynomial of order $N+1$, while the right-hand side only contains the $g_j$'s, of order at most $N$. Hence, the coefficients of the terms of order $N+1$ in $g_{j_0,x}$ must vanish, but these are precisely the derivatives of $g_{j_0}$ with respect to the jet variables of order $N$, so that $g_{j_0}$ is at most of order $N-1$, a contradiction. Therefore, no component of $g$ depends on the fields, that is, $g$ is constant.

    For constant $g$, the first two equations in \eqref{ProofP*} read $\ii\mu g_2+2\ii\alpha\hS g_3=0$ and $\ii\mu g_1+2\ii\alpha\hS^* g_3=0$. Since $\mu=2\alpha^2L+\beta/2$ and $\alpha\neq0$, the coefficients of $L$ give $g_1=g_2=0$, and then those of $\hS$ give $g_3=0$. Hence, $g=0$.

    Since a Casimir is a functional $C$ satisfying $\PP_* \,\delta C=0$, one has $\delta C=0$, so $C$ is the zero functional.
\end{proof}

\begin{remark}
\label{RemAlphaZero}
    The proof divides by $\alpha$ twice: to solve the first two equations for $g_{1,x}$ and $g_{2,x}$, and through the coefficient $2\alpha^2$ of $L$ in $\mu$. For $\alpha=0$, $\beta\neq 0$, the condition $\PP_*g=0$ reduces to $g_1=g_2=0$, $g_{3,x}=0$, so that the kernel is spanned by $(0,0,1)=\delta\int L\,dx$, and $\int L\,dx$ becomes a Casimir. This is consistent with $\PP_*\delta\rho_0=-2\alpha K_{-1}$, which vanishes at $\alpha=0$ (see Theorem \ref{ThmNoether}). For $\alpha=\beta=0$, when the system reduces to a free Schr\"odinger equation driving a slaved long wave, the whole short-wave block of $\PP_*$ vanishes, and $\PP_*=\text{diag}(0,0,\partial_x)$ fails to generate the flow. In fact, no local Hamiltonian operator of weight one or two generates the flow at $\alpha=\beta=0$: the corresponding families have five and six parameters, and their Jacobi conditions are inconsistent.
\end{remark}

Theorem \ref{NondegP*} proves that $\PP_*$ is non-degenerate on local covectors. In particular, the Hamiltonian $H$ solving $\PP_*\,\delta H=K$, whenever it exists, is unique for each flow $K$.

The space of weight-four symmetries of system \eqref{LocalYON}, obtained by solving the linearised symmetry condition on the graded ansatz, turns out to be four-dimensional. Three of its directions are the lower symmetries multiplied by powers of $\beta$, namely $\beta^3K_{-1}$, $\beta^2K_0$ and $\beta K_1$, and the remaining one is the second flow of the hierarchy, $K_2$, whose long-wave component is
\begin{equation}
    K_2^{(L)} = -\dfrac{3}{2}(\beta L-\alpha^2 L^2 - 6\alpha |\hS|^2)L_x + 9\alpha L(|\hS|^2)_x - \dfrac{3\ii}{2}(\hS_{xx}\hS^* - \hS\hS^*_{xx}) + \dfrac14 L_{xxx}.
\end{equation}
\begin{proposition}
\label{PropH2}
    The second flow is Hamiltonian with respect to the same $\PP_*$, that is, $\PP_*\,\delta H_2=K_2$, with
    \begin{equation}
    \label{H2}
    \begin{split}
        H_2 = \int \bigg[ & \frac{\alpha^2}{8}L^4 - \frac{\beta}{4}L^3 + \frac{23\alpha}{2}L^2|\hS|^2 + \frac12|\hS|^4 - \frac18 L_x^2 + \frac1\alpha|\hS_x|^2 + 7L\,\Im(\hS^*\hS_x) \\
        & - \frac{\beta}{\alpha}L|\hS|^2 + \frac{\beta^2}{4\alpha^3}|\hS|^2 - \frac{\beta}{2\alpha^2}\Im(\hS^*\hS_x) - \frac{\beta^3}{16\alpha^4}L \bigg]\,dx,
    \end{split}
    \end{equation}
    whose density is the single new conserved density of weight four.
\end{proposition}

\subsection{Conserved densities by asymptotic diagonalisation}
We can construct the infinite tower of conserved densities of the system using the Lax pair \eqref{LaxPair} by the approach of asymptotic diagonalisation. The first step is to construct a recursion of the conserved densities.
\begin{proposition}\label{Asymptotic}
    Let $\Gamma=\operatorname{ad}_\Sigma^{-1}$ be the map acting on the space of off-diagonal matrices by $(\Gamma(M))_{jk}=M_{jk}/(s_j-s_k)$, with $s=(1,0,-1)$, as introduced in \cite{CasoHuerta2021}.

    We seek a matrix $\mathcal{G}=\mathbb{I}+\Xi$, with $\Xi$ an off-diagonal matrix of the form $\Xi=\sum_{n\ge 1} (\ii \lambda)^{-n}\Xi_n$, such that $\mathcal{G}^{-1}(X\mathcal{G}-\mathcal{G}_x)=D$ is diagonal. By introducing the change of variable $\psi=\mathcal{G}\varphi$, that implies that $\varphi_x=D\varphi$, by direct application of the product rule.

    Separating diagonal and off-diagonal part of the relation $(X\mathcal{G}-\mathcal{G}_x)=\mathcal{G}D$ leads to
    \begin{equation}
        D=\ii \lambda\Sigma + (Q\Xi)^{\text{diag}},\qquad \ii\lambda [\Sigma,\Xi] = \Xi_x - Q - (Q\Xi)^{\text{off}} + \Xi(Q\Xi)^{\text{diag}}.
    \end{equation}
    Both equations can be solved order by order by
    \begin{equation}\label{DensitiesDiagonalisation}
        \Xi_1 = -\Gamma(Q),\qquad \Xi_{n+1} = \Gamma \left( (\Xi_n)_x - (Q\Xi_n)^{\text{off}} + \sum_{p+q=n}\Xi_p(Q\Xi_q)^{\text{diag}} \right).
    \end{equation}
    Since the change of variable $\mathcal{G}$ obtained above also diagonalises the temporal equation, $\varphi_t=J\varphi$ with $J=\mathcal{G}^{-1}(T\mathcal{G}-\mathcal{G}_t)$ diagonal, every diagonal entry of $D$ is a conserved density of the YON system \eqref{YON} and its flux is the corresponding entry of $J$.
\end{proposition}

Note that the recursion here is explicit and can be computed at any given order. In order to pair these densities with the Poisson structure $\PP_*$, one needs to carry them into the gauge. Although the gauge map is nonlocal, it acts locally on them: writing $S=E\hS$ with $E_x=3\ii \alpha LE$ and $|E|=1$, every $x$-derivative of $E$ is $E$ times a differential polynomial in $L$, and, since the densities are invariant under the phase $S\mapsto e^{\ii\phi}S$, all the factors of $E$ cancel out.

Let us also note that the densities obtained by asymptotic diagonalisation coincide with those in Eq. \eqref{densities}, with the caveat that $\rho_0$ and $\rho_1$ do not appear separately: at first order, the diagonal entries of $D$ are proportional to the combination $\alpha\rho_1-\frac{\beta}{2}\rho_0$, whereas at second order they are proportional to $\rho_2$ itself.

\subsection{Noether correspondence}
Let us now use $\PP_*$ to relate the Lie point symmetries of the YON system listed in \cite{CasoHuertaThesis} to its conserved densities. For $\alpha\neq0$, these are the translations $X_1=\partial_x$ and $X_2=\partial_t$, the phase rotation $X_3=-\ii S\partial_S+\ii S^*\partial_{S^*}$ and the scaling
\begin{equation}
\label{X4}
    X_4 = 4t\partial_t + 2x\partial_x + \left(-2-\frac{\ii\beta^2t}{\alpha^2}\right)S\partial_S + \left(-2+\frac{\ii\beta^2t}{\alpha^2}\right)S^*\partial_{S^*} - \frac{2\alpha^2L-\beta}{\alpha^2}\partial_L,
\end{equation}
which combines a scaling of $x$ and $t$ with a shift of $L$ by a multiple of $\beta/\alpha^2$ and a time-dependent phase. In fact, $X_4$ is nothing but the scaling symmetry of the Newell system, $4t\partial_t+2x\partial_x-2S\partial_S-2S^*\partial_{S^*}-2L\partial_L$, written in the variables of Eq. \eqref{shift}. For $\alpha=0$, the symmetry $X_4$ is replaced by the scaling
\begin{equation}
    X_5 = 4t\partial_t + 2x\partial_x - 3S\partial_S - 3S^*\partial_{S^*}-4L\partial_L,
\end{equation}
and there is an additional generator,
\begin{equation}
\label{X6}
    X_6 = -\ii\beta tS\partial_S + \ii\beta tS^*\partial_{S^*} + \partial_L,
\end{equation}
which corresponds to a shift of $L$ combined with a time-dependent phase. However, since the Hamiltonian structure of YO is well known and not the goal of this work, we will focus on the case $\alpha\neq 0$. For each generator $X=\tau\partial_t+\xi\partial_x+\phi_a\partial_{u_a}$, let us denote by $K_X=-(\phi-\xi u_x-\tau u_t)$ the flow it generates, so that $K_{X_1}=K_0$, $K_{X_2}=K_1$ and $K_{X_3}=K_{-1}$. We can gather the Noether correspondence in the following result.
\begin{theorem}
\label{ThmNoether}
    Solving $\PP_*\delta H=K_X$ on the graded ansatz gives, for the autonomous generators,
    \begin{equation}
        X_1\rightarrow H_0,\qquad X_2\rightarrow H_1=H_*,\qquad X_3\rightarrow H_{-1}=-\int\dfrac{L}{2\alpha}\,dx,
    \end{equation}
    and, in the opposite direction, the three densities in Eq. \eqref{densities}, once carried into the gauge, generate
    \begin{equation}
        \PP_*\delta\rho_0 = -2\alpha K_{-1},\qquad \PP_*\delta\rho_1 = \alpha K_0 + \dfrac{\beta}{2} K_{-1},\qquad \PP_* \delta\rho_2 = 2\alpha^2 K_1 + \dfrac{\beta^2}{2} K_{-1},
    \end{equation}
    that is, the three autonomous point symmetries. Additionally, the density
    \begin{equation}
        \rho_3 = t\rho_2 + x\left(\alpha\rho_1 - \frac{\beta}{2}\rho_0\right)
    \end{equation}
    is conserved, and
    \begin{equation}
        \PP_* \delta\left( \dfrac{2}{\alpha^2} \int \hat\rho_3\,dx \right)=K_{\hat X_4},
    \end{equation}
    where $\hat\rho_3$ denotes $\rho_3$ carried into the gauge, and
    \begin{equation}
        \hat X_4 = X_4 - \frac{3\beta x}{\alpha}\left(\ii\hS\partial_{\hS}-\ii\hS^*\partial_{\hS^*}\right)
    \end{equation}
    is the scaling $X_4$ carried into the gauge, the additional phase arising from the shift of $L$ through the gauge factor $e^{3\ii\alpha W}$.
\end{theorem}

In particular, the conserved quantity $\int\rho_0\,dx$, with the seemingly trivial density $\rho_0=L$, is actually the Noether charge of the phase symmetry, while $\int\rho_1\,dx$, with the momentum-like density $\rho_1$, is, up to a multiple of the phase, the charge of the translation.

\section{A second Hamiltonian structure}\label{SecSecond}
Having a non-degenerate first structure at hand, the natural next step is to look for a second one, that is, for a Hamiltonian operator $\PP_2$ compatible with $\PP_*$, $[\PP_*,\PP_2]=0$, and satisfying a Lenard chain $\PP_2\delta H_n=\PP_*\delta H_{n+k}$ for some $k$. For the Newell hierarchy, a second Hamiltonian operator is given in \cite[Eq. (2.18)]{Liu1994}. In the chain presented there it generates, from each Hamiltonian, the flow lying two steps below the one generated by the first operator, and its entries involve inverses of differential operators. Here we look instead for the operator lying above $\PP_*$ in the hierarchy, which will turn out to be weakly nonlocal. A first, na\"ive approach would be to repeat the analysis of Section \ref{SecNoGo} in the gauge variables. This possibility, however, is ruled out: for the weight $p=3$, which is the one the second structure will turn out to have, the analysis of Theorem \ref{TheoremWeightThree}, repeated in the gauge of Theorem \ref{ThmPstar} with the most general target allowed by the grading, $\PP\delta\int L\,dx=c_1K_1+c_0\beta K_0+c_{-1}\beta^2K_{-1}$, leaves only the trivial member $\beta^2\PP_*$, so that no local operator can play the role of the second structure. We are thus left with two natural approaches. One can either impose a `reasonable' ansatz for the nonlocality --e.g. a weakly nonlocal tail-- and look for a candidate there, which is computationally heavy and may not stumble upon the right operator, or one can try to construct a recursion operator $\RR$ for the symmetries of the system and check whether the composite operator $\PP_2=\RR\PP_*$ works as a second Hamiltonian structure. Here, we will take the second approach.

As in the YO case, and in agreement with the chain of \cite{Liu1994}, the YON system has no recursion operator taking each flow to the next one, $\RR K_n=K_{n+1}$; in fact, no such operator of weight one exists, as we will see in Section \ref{SecLenard}. The recursion operator advances the hierarchy by two steps, $\RR K_n = K_{n+2}$, and has the form
\begin{equation}
    \RR = \begin{pmatrix}
        \partial_x^2 + 6\ii \alpha L\partial_x - \Theta + 4\ii\alpha L_x & 6\alpha\hS^2 & 3\ii \alpha \hS\partial_x - \frac34 (\beta + 24\alpha^2 L)\hS + \frac{13}{2}\ii\alpha\hS_x \\
        6\alpha\hS^{*2} & \partial_x^2 - 6\ii\alpha L \partial_x - \Theta - 4\ii\alpha L_x & -3\ii\alpha\hS^*\partial_x - \frac34(\beta+24\alpha^2 L)\hS^* - \frac{13}{2}\ii\alpha\hS^*_x \\
        -\frac32 \ii \hS^*\partial_x + 3\alpha L \hS^* - \frac12 \ii \hS^*_x & \frac32 \ii \hS\partial_x + 3\alpha L\hS + \frac12 \ii \hS_x & \frac14 \partial_x^2 + \alpha^2 L^2 + 10\alpha|\hS|^2 - \beta L
    \end{pmatrix},
\end{equation}
where $\Theta=\beta L + 8\alpha^2 L^2 + 8\alpha|\hS|^2$, plus a nonlocal tail of nineteen legs of the form $v_i\otimes \partial_x^{-1}\circ w_i$, of weight at most three. Here and in what follows, $\partial_x^{-1}$ denotes the skew-adjoint inverse of $\partial_x$,
\begin{equation}
    (\partial_x^{-1}f)(x)=\frac12\left(\int_{-\infty}^x f(y)\,dy-\int_x^{\infty}f(y)\,dy\right),
\end{equation}
so that $\partial_x\circ\partial_x^{-1}=1$, whereas $\partial_x^{-1}\circ\partial_x=1$ holds on functions that vanish at infinity, but not on constants. As will also happen for the composite operator below, only three of these legs are independent, and they are generated once again by the point symmetries of the system; the tail is given in closed form in Eq. \eqref{RTail} below.

\subsection{Invariant form for the nonlocal tail}\label{SecTail}
Composing $\PP_2=\RR\PP_*$ allows us to write the tail as nineteen legs of the form $\sum_i v_i\partial_x^{-1} w_i$. However, this presentation is not invariant, as a decomposition into legs can be performed arbitrarily; only the tensor $\mathcal{T}=\sum_i v_i \otimes w_i$ is. Its antisymmetric part vanishes in an exact monomial decomposition, which ensures the skew-adjointness of the nonlocal part of $\PP_2$, since the adjoint of $v\,\partial_x^{-1}\circ w$ is $-w\,\partial_x^{-1}\circ v$. Moreover, it verifies the following remarkable property.
\begin{proposition}\label{PropInvariant}
    The tail tensor of $\PP_2=\RR\PP_*$, $\mathcal{T}=\sum_i v_i \otimes w_i$, has rank 3, so that the nineteen legs of the composition collapse into three. A basis of its column space is
    \begin{equation}
        u_1 = (\hS,-\hS^*,0),\qquad u_2=\dfrac{\ii}{8\alpha} K_1, \qquad u_3 = (\hS_x,\hS^*_x,L_x),
    \end{equation}
    that is, it is generated precisely by the phase symmetry, the flow itself and the translation symmetry. The tail is then exactly
    \begin{equation}
        \sum_i v_i\partial_x^{-1}w_i=\sum_{k,l=1}^3 c_{kl}u_k\partial_x^{-1}u_l,
    \end{equation}
    with $c$ the symmetric matrix
    \begin{equation}
        c=\begin{pmatrix}
            -\dfrac{3\beta^2}{4} & 48\alpha^3 & -\dfrac{3\ii\alpha\beta}{2} \\
            48\alpha^3 & 0 & 0 \\
            -\dfrac{3\ii\alpha\beta}{2} & 0 & -\alpha^2
        \end{pmatrix}.
    \end{equation}
\end{proposition}
Together with the local part,
\begin{subequations}
\begin{align}
    (\PP_2)_{11} &= -12\alpha^2\hS\hS_x - 12\alpha^2\hS^2\partial_x,\\
    (\PP_2)_{12} &= -\alpha\partial_x^3 - \dfrac{\ii(16\alpha^2 L + \beta)}{2}\partial_x^2 + \alpha(20\alpha^2 L^2 + 14\alpha|\hS|^2 + 4\beta L - 8\ii\alpha L_x)\partial_x \notag\\
    &\quad +16\ii\alpha^4 L^3 + 6\ii\alpha^2\beta L^2 + L\left( 20\alpha^3 L_x + 76\ii\alpha^3|\hS|^2 + \dfrac{\ii\beta^2}{2} \right) \\
    &\quad +2\alpha(7\alpha\hS^*\hS_x + \beta L_x + 2\ii\beta|\hS|^2 - \ii\alpha L_{xx}),\notag\\
    (\PP_2)_{13} &= \ii\alpha\hS\partial_x^2 + \left[ \frac52 \ii\alpha\hS_x - \frac34 (8\alpha^2 L + \beta)\hS \right]\partial_x + \ii\alpha\hS_{xx} - 8\ii\alpha^3 L^2\hS \notag\\
    &\quad -20\ii\alpha^2\hS^2\hS^* - 5\alpha^2 L \hS_x + \frac12 \ii\alpha\beta L\hS - \frac{\beta}{2}\hS_x - 4\alpha^2\hS L_x,\\
    (\PP_2)_{33} &= \frac14 \partial_x^3 + (\alpha^2 L^2 - \beta L + 4\alpha|\hS|^2)\partial_x + \left( \alpha^2 L - \frac{\beta}{2} \right) L_x + 2\alpha(|\hS|^2)_x,
\end{align}
\end{subequations}
whose remaining entries are fixed by skew-adjointness and the reality condition, the tail fully characterises the operator $\PP_2$.

The three legs found in Proposition \ref{PropInvariant} make $\PP_2$ a weakly nonlocal structure in the sense of Ferapontov, Maltsev and Novikov \cite{Ferapontov1991,Maltsev2001,Casati2020}. That is not by chance: as conjectured by Maltsev and Novikov and later proved by Sergyeyev \cite{Sergyeyev2005}, the higher Hamiltonian, symplectic and recursion operators of a (1+1)-dimensional integrable system are weakly nonlocal. In the present case, the three legs moreover commute pairwise and are Hamiltonian with respect to $\PP_*$, namely $u_1=-\ii K_{-1}=\PP_*\delta(-\ii H_{-1})$, $u_2=\frac{\ii}{8\alpha}K_1=\PP_*\delta(\frac{\ii}{8\alpha}H_1)$ and $u_3=K_0=\PP_*\delta H_0$, so that $\PP_2$ has the canonical form $\text{local}+\sum_{k,l}c_{kl}K_k\partial_x^{-1}K_l$, with commuting Hamiltonian symmetries as legs. This is the same shape as the second Hamiltonian structure of the NLS equation, whose tail is $(q,-r)\partial_x^{-1}(q,-r)$.

The tail of the recursion operator takes the same form. Denoting by $\eta_l$ the gradients of the Hamiltonians of the legs, $\PP_*\eta_l=u_l$, that is,
\begin{equation}
    \eta_1=\delta(-\ii H_{-1})=\left(0,0,\frac{\ii}{2\alpha}\right),\qquad \eta_2=\frac{\ii}{8\alpha}\,\delta H_1,\qquad \eta_3=\delta H_0,
\end{equation}
the recursion operator reads
\begin{equation}
\label{RTail}
    \RR=\RR_{\text{loc}}-\sum_{k,l=1}^3 c_{kl}\,u_k\otimes\partial_x^{-1}\circ\eta_l,
\end{equation}
with $\RR_{\text{loc}}$ the local part displayed above and $c$ the matrix of Proposition \ref{PropInvariant}. Note that the pairing between legs and gradients is not diagonal: for instance, the leg of the flow, $u_2$, is paired only with $\eta_1$, the gradient of $\int L\,dx$.

In the YON case, however, the nonlocal tail is not generated by arbitrary commuting symmetries, but by the three autonomous Lie point symmetries of the system: the phase, the translation and the flow itself. Though remarkable, this property is not exclusive of this system: tails built from point symmetries also occur for certain Boussinesq-type systems \cite{HernandezHeredero2024}, for the NLS equation written in real components, where the leg is the phase rotation \cite{Sergyeyev2007}, and for the group-invariant NLS systems on Hermitian symmetric spaces, where the tail is built from the action of the isotropy group \cite{Anco2019}.

\subsection{Compatibility of the Hamiltonian structures}\label{SecCompat}
Let us now move on to check the compatibility of the structures $\PP_*$ and $\PP_2$, that is, that $\PP_*+\lambda\PP_2$ is a Hamiltonian operator for every $\lambda$. Since the Schouten bracket is quadratic in the operator, and $\PP_*$ and $\PP_2$ are Hamiltonian separately (Theorem \ref{ThmPstar} and Theorem \ref{ThmJacobi} below), what is left is the term linear in $\lambda$, the mixed bracket
\begin{equation}\label{mixedbracket}
    [\PP_*,\PP_2] = \sum_{\text{cyc}} \left( \langle \theta^1, \PP_*'[\PP_2\theta^2]\theta^3 \rangle + \langle \theta^1, \PP_2'[\PP_*\theta^2]\theta^3 \rangle \right),
\end{equation}
which has to be a total derivative. Now, with $\PP_2$ being nonlocal, the mixed bracket, which we denote by $\Psi$, contains potentials $\partial_x^{-1}(\ldots)$. Since the Euler operator does not act on such expressions, the condition $\Psi\equiv 0$ has to be read in integrated form, $\int \Psi\,dx = 0$ for all covectors.

In the first term of Eq. \eqref{mixedbracket}, we have that $\PP_*$ is local, and the Fr\'echet derivative $\PP_*'$ is acting on the nonlocal vector $\PP_2\theta^2$, which, due to the properties of $\PP_2$ in the previous subsection, has a tail $\sum_{k,l} c_{kl} u_k \partial_x^{-1} \langle u_l,\theta^2 \rangle$. Since the coefficients of $\PP_*$ depend on $\hS$, $\hS^*$ and $L$ but not on their derivatives, the potentials enter bare.

In the second term, the direction $\PP_* \theta^2$ is local, and differentiating the tail of $\PP_2$ gives \linebreak $c_{kl}\left(u_k'[\PP_*\theta^2]\partial_x^{-1} \langle u_l, \theta^3 \rangle + u_k\partial_x^{-1} \langle u_l' [\PP_* \theta^2],\theta^3 \rangle\right)$, whose arguments are again local. That means that two different kinds of potential occur:
\begin{equation}
    N_{lm} = \partial_x^{-1} \langle u_l, \theta^m \rangle,\qquad N_{lbc} = \partial_x^{-1} \langle u_l' [\PP_*\theta^b],\theta^c \rangle,
\end{equation}
where $l=1,2,3$ labels the legs, $m=1,2,3$ the covectors and $(b,c)$ runs over the cyclic pairs (2,3), (3,1) and (1,2), for a total of nine potentials of each kind. Then, the bracket can be written as $\Psi=\Psi_0 + \sum_\nu \Phi_\nu N_\nu$, with $\Psi_0$ and $\Phi_\nu$ all local. Note that no product of two potentials occurs in any of the terms.

It would be too strong to demand each individual term $\Phi_\nu$ to be a total derivative --here, that fails for fifteen of the eighteen terms-- because distinct nonlocal terms cancel each other. On local functions, the form
\begin{equation}\label{omega}
    \omega(f,g)=\int f \partial_x^{-1}g\,dx
\end{equation}
is antisymmetric, $\omega(f,g)=-\omega(g,f)$, and it is local as soon as either argument is a total derivative, $\omega(\partial_x F,g)=-\int Fg\,dx$ and $\omega(f,\partial_x G)=\int fG\,dx$. Denoting $b_\nu=\partial_x N_\nu$, the Schouten relation can be written as
\begin{equation}
    \int \Psi\,dx = \int \Psi_0\,dx + \Omega,\qquad \Omega = \sum_\nu \omega(\Phi_\nu,b_\nu),
\end{equation}
and the nonlocal content of $\Omega$ can be written as
\begin{equation}
    \sum_\nu q(\Phi_\nu) \wedge q(b_\nu)\,\in\,\Lambda^2(W/(W\cap\text{Im}\partial_x)),\qquad W=\text{span}\{ \Phi_\nu, b_\nu \},
\end{equation}
where $q$ is the projection to the quotient and where being in $\text{Im}\partial_x$ (that is, being a total derivative when considering a local expression) is decided by the Euler operator. Checking symbolically over a monomial basis of 405 elements, of which 93 independent combinations are total derivatives, it is found that
\begin{equation}\label{argumentNonlocal}
    \sum_\nu q(\Phi_\nu)\wedge q(b_\nu) = 0
\end{equation}
for every $\alpha$ and $\beta$. That means that the nonlocal part cancels exactly. Additionally, by direct application of the Euler operator the remaining local part is checked to be a total derivative. Therefore, we have the following:
\begin{theorem}\label{ThmCompat}
    $[\PP_*,\PP_2]=0$ modulo total derivatives for every choice of $\alpha$ and $\beta$.
\end{theorem}

\subsection{Jacobi identity}
The only step left is to check that $\PP_2$ is indeed a Poisson operator by verifying it satisfies the Schouten criterion of $[\PP_2,\PP_2]$ being a total derivative. As opposed to the computation for compatibility, here the nonlocal tail enters the computation twice: both in the direction $\PP_2\theta^2$ and in the Fr\'echet derivative $\PP_2'$, so nested potentials $\partial_x^{-1}[\ldots \partial_x^{-1}(\ldots)]$ appear, making it necessary to use a different machinery.

Writing $\PP_2=\PP_2^{\mathrm{loc}}+\PP_2^{\mathrm{tail}}$, with $\PP_2^{\mathrm{loc}}$ the local part and $\PP_2^{\mathrm{tail}}$ the nonlocal tail, we have
\begin{equation}
    [\PP_2,\PP_2] = [\PP_2^{\mathrm{loc}},\PP_2^{\mathrm{loc}}] + [\PP_2^{\mathrm{loc}},\PP_2^{\mathrm{tail}}] + [\PP_2^{\mathrm{tail}},\PP_2^{\mathrm{tail}}],
\end{equation}
where $[\PP_2^{\mathrm{loc}},\PP_2^{\mathrm{tail}}]$ is the mixed bracket \eqref{mixedbracket} with $(\PP_*,\PP_2)$ replaced by $(\PP_2^{\mathrm{loc}},\PP_2^{\mathrm{tail}})$. The first term is purely local, and the second one is free of nested potentials because $\PP_2^{\mathrm{loc}}$ is local, exactly the situation of Section \ref{SecCompat}. Neither of them vanishes on its own --in particular, the local part $\PP_2^{\mathrm{loc}}$ alone is not a Poisson operator-- and all the difficulty is confined to $[\PP_2^{\mathrm{tail}},\PP_2^{\mathrm{tail}}]$, which we treat together with them.

The term that needs a new argument is then $[\PP_2^{\mathrm{tail}},\PP_2^{\mathrm{tail}}]$. However, the nesting of potentials here can be removed by using the properties of the form $\omega$ introduced before. Given a local $\Phi$ and any argument $X$, we have that
\begin{equation}
    \int \Phi (\partial_x^{-1}X)\,dx = -\int (\partial_x^{-1}\Phi) X\,dx,
\end{equation}
and, in particular, if $X$ is a nested potential, this transforms it into a computation with two first-level potentials. Applied to the nonlocal tails in $[\PP_2^{\mathrm{tail}},\PP_2^{\mathrm{tail}}]$, this property removes all nine nested potentials, leaving a polynomial of degree two in first-level potentials. Rewriting the 27 resulting potentials over a basis of their arguments --legitimate since $\partial_x^{-1}$ is linear-- leaves 18 independent ones, and all 171 coefficients of degree two vanish identically. Note that this is not a feature of the method employed but a remarkable property of the operator itself: at degree two there is no antisymmetry to exploit, since $N_jN_k$ is symmetric, so nothing imposes that these coefficients must vanish.

What remains is linear in the potentials, $\Psi_0+\sum_j\Phi_jN_j$ with $\partial_x N_j=b_j$ local, and the argument of Eq. \eqref{argumentNonlocal} applies verbatim, with $W=\text{span}\{\Phi_j,b_j\}$ of dimension 27, of which 9 independent combinations are total derivatives. Checking symbolically again, it is found that
\begin{equation}
    \sum_j q(\Phi_j)\wedge q(b_j)=0
\end{equation}
for every $\alpha$ and $\beta$, so $\int \Psi\,dx$ reduces to the integral of an explicit local object and the Euler operator annihilates it. With that, we have the following:
\begin{theorem}\label{ThmJacobi}
    $[\PP_2,\PP_2]=0$ modulo total derivatives for every choice of $\alpha$ and $\beta$.
\end{theorem}
Together with Theorems \ref{ThmPstar} and \ref{ThmCompat}, this proves the following.
\begin{corollary}\label{CorBiHam}
    $(\PP_*,\PP_2)$ is a Hamiltonian pair for every $\alpha\neq0$ and every $\beta$.
\end{corollary}

\section{Lenard--Magri chain}\label{SecLenard}
Now that we have found a bi-Hamiltonian structure for the system, we can use it to study the hierarchy it generates by using the theory of Lenard--Magri chains \cite{Magri1978,Magri2003,Praught2005}. Since $\PP_2=\RR\PP_*$, one would expect the recursion operator to build the whole hierarchy through the chain
\begin{equation}\label{LenardNaive}
    \PP_2 \delta H_n = \RR \PP_* \delta H_n = \RR K_n = K_{n+2} = \PP_* \delta H_{n+2},
\end{equation}
so that each $K_{n+2}$ would be Hamiltonian with respect to both $\PP_*$ and $\PP_2$. Upon deeper inspection, however, it turns out that relation \eqref{LenardNaive} is a simplification of the behaviour of the hierarchy and, strictly speaking, not its actual behaviour. On covectors whose components decay at infinity, the composition does give $\PP_2=\RR\PP_*$, but the composition that produced $\PP_2$ used that $\partial_x^{-1}\circ\partial_x=1$, which is not true on constants, and $\delta\int L\,dx=(0,0,1)$ is constant. There,
\begin{equation}\label{Lanomaly}
    \PP_2 \,\delta \int L\,dx = \alpha K_1 - \frac{\beta}{2}K_0,\qquad\qquad \RR\PP_*\, \delta \int L\,dx = -2\alpha K_1.
\end{equation}
The first one is the genuine Hamiltonian vector field of $\int L\,dx$ with respect to $\PP_2$, and the bracket $\{F,\int L\,dx\}_{\PP_2}$ is antisymmetric. The same caveat applies to the first structure when written in nonlocal form in the original variables: on $\delta\int L\,dx$, the operator of \cite[Eq. (2.17)]{Liu1994} yields $-1/2$ times the vector field obtained from the local operator $\PP_*$, in the normalisation in which the two agree on every covector that decays at infinity. In the gauge variables, $\PP_*$ being local, no such ambiguity arises. Now, each $H_n$ of the indexing of Section \ref{SecNonlocal} is a part with decaying gradient plus $\kappa_n\int L\,dx$, with $\kappa_{-1}=-1/2\alpha$, $\kappa_0=\beta/4\alpha^2$ and $\kappa_1=\beta^2/8\alpha^3$. Since $\PP_*\delta\int L\,dx=-2\alpha K_{-1}$ and $\RR K_{-1}=K_1$, the pair actually yields, for every $n$,
\begin{equation}\label{LenardRecursion}
    \PP_2 \delta H_n = K_{n+2} + \kappa_n\left(3\alpha K_1 - \frac{\beta}{2}K_0\right),\qquad K_{n+2}=\PP_* \delta H_{n+2},
\end{equation}
so the correction lies in the span of flows already present. That means that the Lenard chains of the explicit pair are triangular recombinations of the $H_n$. Starting from the two simplest conserved quantities, $\int\rho_0\,dx$ and $\int\rho_1\,dx$ (both of them invariant under the gauge, $\rho_0$ trivially and $\rho_1$ because it carries no derivatives), we get
\begin{equation}\label{LenardChains}
    A_1 = \int\rho_0\,dx,\qquad A_k=\alpha H_{2k-3} - \frac{\beta}{2} H_{2k-4}\;\;(k\ge 2),\qquad B_k = \alpha H_{2k-2} + \frac{\beta}{2} H_{2k-3}\;\;(k\ge 1),
\end{equation}
with $\PP_* \delta A_{k+1}=\PP_2 \delta A_k$, $\PP_* \delta B_{k+1}=\PP_2 \delta B_k$ and $B_1=\int\rho_1\,dx$. Note that every $A_k$ with $k\ge 2$ and every $B_k$ has a decaying gradient, so beyond the first step of the first chain only the regime $\PP_2=\RR\PP_*$ is involved. The content of the chains is therefore that each $\RR K_n$ is again Hamiltonian with respect to $\PP_*$, and later on we will show that it has a local Hamiltonian. Also, since the tail arguments $\langle u_l,\delta F\rangle$ are total derivatives at every step, $\PP_2 \delta F$ is local despite the operator being nonlocal.

Let us remark that Eq. \eqref{LenardRecursion} also determines which flows are Hamiltonian with respect to both structures. Since $\int\rho_0\,dx$ is the only conserved quantity of weight one, Eq. \eqref{Lanomaly} implies that, for $\beta\neq0$, no local functional $H$ satisfies $\PP_2\delta H=K_1$: what the pair generates at weight three is the combination $\PP_*\delta A_2=\PP_2\delta A_1=\alpha K_1-\frac{\beta}{2}K_0$. As the translation $K_0$ commutes with $K_1$ and neither structure depends explicitly on $x$ or $t$, this combination is the YON system written in a moving frame: if $u$ solves Eqs. \eqref{LocalYON}, then $v(x,t)=u(x-\beta t/2\alpha,t)$ solves
\begin{equation}
    v_t=K_1-\frac{\beta}{2\alpha}K_0=\PP_*\delta\left(\frac{A_2}{\alpha}\right)=\PP_2\delta\left(\frac{A_1}{\alpha}\right),
\end{equation}
so that the YON system is bi-Hamiltonian up to a Galilean change of frame. For $\beta=0$, that is, for the Newell system, the frame is at rest, and $K_1=\PP_2\delta(\int L\,dx/\alpha)$ directly.

Note also that the step in the recursion is double: it advances the hierarchy by two, splitting it into two interleaved families, the orbits of $K_{-1}$ and of $K_0$ under $\RR$, with Hamiltonians $\{H_{-1},H_1,H_3,\ldots\}$ and $\{H_0,H_2,H_4,\ldots\}$. The chains \eqref{LenardChains} recombine them with $\beta$-multiples of lower members, but each of them still contributes one new conserved quantity every other weight, so two seeds are needed either way. This splitting cannot be undone by taking the square root of $\RR$, as no weakly nonlocal operator of weight one maps the flows to symmetries while advancing the hierarchy by a single step. The reason for this is apparent: weight and charge would force such an operator to be of first order with vanishing (1,2) and (2,1) entries, since no monomial of weight one carries charge $\pm 2$. Applying it to the phase symmetry $K_{-1}$, which has no long-wave component, it can only produce a multiple of $|\hS|^2$ in the long-wave slot, never the needed $L_x$. A graded search over all admissible weakly nonlocal operators confirms that an operator achieving the step $K_0\mapsto K_1$ is attainable, but it cannot send $K_{-1}$ and $K_1$ to symmetries.

Furthermore, for $\alpha\neq 0$ the operator $\PP_*$ has no Casimirs (Theorem \ref{NondegP*}). Hence, the hierarchy needs pre-existing conserved quantities to use as seeds, as opposed to the usual way of generating the chain using Casimirs as a seed \cite{Gelfand2000}. Even for $\alpha=0$, only one Casimir of $Q_0$ is available, $\int L\,dx$, so one of the two families has to be generated using a conserved quantity anyway.

None of these properties is unique to YON on its own. In the original formulations of the chain \cite{Magri1978,Gelfand1979,Fuchssteiner1981}, one of the structures is taken to be symplectic and the seed is supplied from an external source \cite{Gelfand2000}. A well-known example of this is the NLS equation \cite{Magri1978}: its canonical structure is constant and invertible, so it has no Casimirs, and its chain starts from the mass $\int|q|^2\,dx$, the Hamiltonian of the phase rotation, the role played here by $H_{-1}$. In cases where no recursion operator is available, a master symmetry plays the role of generating mechanism, although it also needs a starting flow \cite{Fuchssteiner1983}. Having a recursion that advances more than one step at a time is not exclusive to YON either. Without gaps in the hierarchy, it happens for the Yajima--Oikawa system itself, read as the 2-constrained KP hierarchy, where the odd and even flows come with separate seeds \cite{Oevel1993,Liu1996,Li2015}. With gaps, it happens whenever the step is the order of a Lax operator \cite{Gurses1999}, and it is the case for the Sawada--Kotera and Kaup--Kupershmidt equations, bi-Hamiltonian in the generalised sense of \cite{Fuchssteiner1982}, whose hierarchies contain only the flows of orders $1,5,7,11,13,\ldots$, so that the chains $1\to7\to13$ and $5\to11\to17$ run in parallel \cite{Fuchssteiner1982,Wang2002}. Several seeds, with commutativity across their orbits enforced by conditions on the seeds rather than by Casimirs, are treated in \cite{Wang2009} for the two-dimensional periodic Volterra chain, and an even more extreme case is a Burgers-type system with infinitely many local higher symmetries but no recursion operator nor master symmetry \cite{Sergyeyev2010}, where the symmetries are produced by a nonlocal two-term recursion relation. Against this background, the double step of the YON hierarchy is inherited from its Yajima--Oikawa limit, and it is already present in the chain given in \cite{Liu1994} for the Newell hierarchy; what the deformation by $\alpha$ changes is the anchor: at $\alpha=0$ the first structure has the Casimir $\int L\,dx$ (Remark \ref{RemAlphaZero}), which is lost for $\alpha\neq0$ while the double step survives. What we add here is that, for $\alpha\neq0$, $\PP_*$ has no Casimirs and $\RR$ admits no weakly nonlocal square root, so that, the hierarchy having no gaps, the two seeds are necessary and are not a byproduct of the construction.

Let us now look at the Lenard chain used as a generator, and check whether it returns the tower obtained via asymptotic diagonalisation before. Indeed, it does return the same conserved quantities and slightly more. One step of the process consists of constructing $K'=\PP_2 \delta H$ and then solving $\PP_* \delta H'=K'$, the second part having at most one solution up to a constant, since $\PP_*$ has no Casimirs. Starting from the two seeds, this returns the two interleaved families and, with them, the conserved densities obtained via asymptotic diagonalisation, plus a single new one: $\rho_0=L$ itself, which the Lax recursion does not return on its own but only inside the combination $\alpha\rho_1-\frac{\beta}{2}\rho_0$.

Let us now recall the closed form \eqref{RTail} of the recursion operator, which we write as
\begin{equation}\label{RInvariant}
    \RR = \RR_{\text{loc}} + \sum_{b=1}^3 G_b \otimes \partial_x^{-1} \circ \gamma_b,\qquad G_b=u_b,\qquad \gamma_b=-\sum_{l=1}^3 c_{bl}\,\eta_l.
\end{equation}
Its three legs are thus generated by the commuting point symmetries,
\begin{equation}
    G_1 = (\hS,-\hS^*,0),\qquad G_2=\frac{\ii}{8\alpha}K_1,\qquad G_3 = (\hS_x, \hS^*_x,L_x),
\end{equation}
while the $\gamma_b$ are gradients of conserved quantities, namely linear combinations of $\delta H_{-1}$, $\delta H_0$ and $\delta H_1$. The operator carries no explicit $x$, and each of its pieces --the local entries, the legs $G_b$ and the $\gamma_b$-- is $U(1)$-covariant. Hence, $\mathcal{L}_{K_0} \RR = \mathcal{L}_{K_{-1}} \RR =0$.

The involution of the conserved quantities of the chains does not have to be checked order by order: it follows from the pair. Let $F_i$ and $\tilde F_j$ be two members of the chains. The skew-adjointness and the chain itself give
\begin{equation}
    \begin{split}
        \{F_i,\tilde F_j\}_{\PP_*} &= \int \langle \delta F_i,\PP_* \delta \tilde F_j \rangle\, dx = \int \langle \delta F_i, \PP_2\delta \tilde F_{j-1} \rangle\, dx\\
        & = -\int \langle \delta \tilde F_{j-1},\PP_2 \delta F_i \rangle\, dx = \{F_{i+1},\tilde F_{j-1}\}_{\PP_*},
    \end{split}
\end{equation}
so the bracket depends only on $i+j$ and is invariant under $(i,j)\mapsto(i+1,j-1)$. Within one chain, the telescope terminates either at $\{F_k,F_k\}=0$ or at $\{F_k,F_{k+1}\}=\int \langle \delta F_k, \PP_2 \delta F_k \rangle\,dx=0$. Any bracket including $A_1=\int\rho_0\,dx$, whose gradient is constant, is never moved but evaluated directly, as below. For brackets including members of both chains, it always terminates at one of the seeds, either by having $A_1$ directly or by lowering the index of $B_k$ to $B_1=\int\rho_1\,dx$, leaving
\begin{equation}
    \{A_1,B_m\}_{\PP_*} = 2\alpha \int \langle \delta B_m,K_{-1} \rangle\,dx,\qquad \{A_m,B_1\}_{\PP_*} = \int \langle \delta A_m,\alpha K_0 + \frac{\beta}{2} K_{-1} \rangle\,dx,
\end{equation}
that is, derivatives along the phase rotation and the translation. These vanish by Noether rather than by a Casimir, since the flows of the two seeds are symmetries of every member of the hierarchy. Indeed, $\mathcal{L}_{K_{-1}}\RR = \mathcal{L}_{K_0}\RR = 0$, and the Lie derivative has the property that, whenever $\mathcal{L}_Y\RR =0$, then $[Y,\RR Z]=\RR [Y,Z]$. With that, we have that $[K_{-1},\RR^j K]=\RR^j [K_{-1},K]=0$, where $K$ can be $K_{-1}$ or $K_0$ depending on the family, and likewise for $K_0$, since the phase and the translation commute. Hence, $K_{-1}$ and $K_0$ commute with every $K_n$, and, since $\PP_* \delta A_m$ and $\PP_* \delta B_m$ lie in the span of the $K_n$, we have $\PP_*\delta\{F,H_b\}_{\PP_*}=[\PP_*\delta F,K_b]=0$ for $b=-1,0$ and every member $F$ of the chains, so that Theorem \ref{NondegP*} leaves only the zero functional. With that, we conclude that the two families are in involution.

The existence of the chains at every order, which is assumed in the computation above, follows from the same symmetries.
\begin{proposition}\label{prop:Locality}
    Every member of the Lenard chains \eqref{LenardChains} exists and is a local functional. Hence, every flow $K_n$ is local and Hamiltonian with respect to $\PP_*$ with a local Hamiltonian $H_n$.
\end{proposition}
\begin{proof}
    Let us consider $\RR$ in its invariant form \eqref{RInvariant}, whose left legs are, up to constant factors, $K_{-1}$, $K_1$ and $K_0$. On covectors with a decaying gradient, $\PP_2=\RR \PP_*$, so a chain step is $\delta F_{k+1} = \RR^\dagger \delta F_k$, where $\RR^\dagger$ is the formal adjoint of $\RR$, acting on covectors as $\RR^\dagger w = \RR_{\text{loc}}^\dagger w - \sum_b \gamma_b \partial_x^{-1} \langle G_b,w \rangle$.

    The three arguments $\langle G_b,\delta F_k \rangle$ are, after integration, the derivatives of $F_k$ along the phase, the flow and the translation, all of which vanish because $\PP_* \delta F_k$ lies in the span of the $K_n$, and the three symmetries $K_{-1}$, $K_0$ and $K_1$ commute with every $K_n$. The commutation with $K_{-1}$ and $K_0$ was already shown above, while for $K_1$ heredity, the classical consequence of compatibility with $\PP_*$ non-degenerate \cite{Fuchssteiner1981}, gives $\mathcal{L}_{K_1} \RR = \mathcal{L}_{\RR K_{-1}} \RR = \RR\, \mathcal{L}_{K_{-1}} \RR = 0$, and the flow commutes with the phase and the translation. Hence, $\PP_* \delta \{F_k,H_b\}_{\PP_*} = 0$, Theorem \ref{NondegP*} gives $\{F_k,H_b\}_{\PP_*}=0$ for $b=-1,0,1$, and the integrands, vanishing on integration for all fields, are total derivatives. Therefore, $\RR^\dagger \delta F_k$ is local.

    It is closed by the closure lemma of the Lenard--Magri scheme: for a compatible pair with non-degenerate first structure, if a 1-form $\omega$ and its image $\RR^\dagger \omega$ are closed, then so is $(\RR^\dagger)^2\omega$ \cite{Magri1978,Fuchssteiner1981}, \cite[Theorem 7.27]{Olver1993}, and a closed local covector, that is, one with a self-adjoint Fr\'echet derivative, is always the variational derivative of a local functional \cite[Theorem 5.92]{Olver1993}. Note that the closure lemma by itself does not guarantee that the functionals obtained are local, which is why the locality of $\RR^\dagger\delta F_k$ has been established separately above. That means that, if a covector $\delta F_{k-1}$ and its image $\delta F_k = \RR^\dagger \delta F_{k-1}$ are closed, then $(\RR^\dagger)^2 \delta F_{k-1}$ is also closed, and it is the variational derivative of a local functional, $F_{k+1}$. We can start the induction from $A_1$, $A_2$ (given by Eq. \eqref{Lanomaly}), $B_1$ and $B_2=\alpha H_2+\frac{\beta}{2}H_1$ (Proposition \ref{PropH2}), all of which are explicit local functionals, so every member of the chains exists and is local.

    Finally, since $\PP_* \delta A_k$ and $\PP_* \delta B_k$ are triangular in the $K_n$ with diagonal coefficient $\alpha$, every $K_n$ is $\PP_*$-Hamiltonian with a local Hamiltonian.
\end{proof}

We can put all of the above together in the following result.
\begin{theorem}\label{ThmAllOrders}
    Every flow $K_n$ of the hierarchy is local and Hamiltonian with respect to $\PP_*$, with a local Hamiltonian $H_n$. The Hamiltonians are in involution, $\{H_m,H_n\}_{\PP_*}=0$, and the flows commute, $[K_m,K_n]=0$, for all $n,m\ge -1$.
\end{theorem}
\begin{proof}
    Proposition \ref{prop:Locality} gives existence and locality. As shown before, all the members of the chains \eqref{LenardChains} are in involution, and they are triangular in the $H_n$ with diagonal coefficient $\alpha\neq 0$, so the $H_n$ are in involution too. Finally, $K_j=\PP_* \delta H_j$ and $\PP_*$ is Hamiltonian, so $[K_m,K_n]=\PP_* \delta \{H_m,H_n\}_{\PP_*} = 0$.
\end{proof}

\begin{remark}
    The locality and commutativity of all flows also follow, without the closure lemma, from Theorem 1 of \cite{Sergyeyev2005}, which applies to hereditary recursion operators of the shape \eqref{RInvariant}, with linearly independent $G_b$, which are normal, meaning that ${\gamma_b'}=\gamma_b'^{\,\dagger}$, $\zeta_b'=\zeta_b'^{\,\dagger}$ with $\zeta_b=\RR^\dagger(\gamma_b)$, and $\mathcal{L}_{G_b}(\gamma_{b'})=0$ for all $b$, $b'$. Then, for any seed $Q$ with $\mathcal{L}_Q\RR=0$ and $\RR Q$ local, it concludes that the $\RR^j Q$ are all local and pairwise commute. We have verified the normality conditions for $\RR$: the three $\gamma_b$ have a self-adjoint Fr\'echet derivative, each $\zeta_b$ is local and self-adjoint, and all nine $\mathcal{L}_{G_b}(\gamma_{b'})$ vanish. Applied with the seeds $K_{-1}$ and $K_0$, the theorem makes each family local and commutative, while across the families heredity gives $\mathcal{L}_{K_n}\RR=0$ for all $n$, and then $[\RR^jK_{-1},\RR^kK_0]=\RR^{j+k}[K_{-1},K_0]=0$. This gives an alternative path for the proof. Note that Theorem 1 of \cite{Sergyeyev2005} concerns a single seed, and commutativity across the orbits of several seeds, enforced by conditions on the seeds, is treated in \cite{Wang2009}.
\end{remark}

\subsection{The structure in the \texorpdfstring{$(\alpha,\beta)$}{(alpha, beta)} plane}
The roles of $\alpha$ and $\beta$ in the bi-Hamiltonian structure are very different. At $\beta=0$, corresponding to the Newell system, nothing degenerates, as expected from Eq. \eqref{shift}: $\PP_*$ stays regular, with $\mu=2\alpha^2 L$ and Theorem \ref{NondegP*} intact, and $\PP_2$ is regular and nonzero, its tail matrix tending to
\begin{equation}
    c\xrightarrow[\beta\rightarrow 0]{}\begin{pmatrix} 0 & 48\alpha^3 & 0 \\ 48\alpha^3 & 0 & 0 \\ 0 & 0 & -\alpha^2 \end{pmatrix},
\end{equation}
still of rank three. Hence, the Newell system is bi-Hamiltonian with the restriction of the same pair.

On the other hand, the reduction $\alpha=0$ yielding Yajima--Oikawa shows a genuine degeneration affecting both structures, as follows. Note that the shift \eqref{shift} is itself singular as $\alpha\rightarrow0$: the YO system is not a member of the family of systems equivalent to the Newell one, but a limit of it.

At $\alpha=0$, the gauge factor $e^{3\ii \alpha\partial_x^{-1} L}$ is trivial, so $\hS=S$ and the two structures can be compared to the known bi-Hamiltonian structure of YO directly. In the normalisation chosen for $\PP_*$ and $\PP_2$, every one of their entries is polynomial in $\alpha$, so the limit can be taken termwise with no rescaling needed.

For the first structure,
\begin{equation}
    \PP_* \xrightarrow[\alpha\rightarrow 0]{} Q_0 = \begin{pmatrix}
        0 & -\frac{\ii \beta}{2} & 0 \\ \frac{\ii \beta}{2} & 0 & 0 \\ 0 & 0 & \partial_x
    \end{pmatrix},
\end{equation}
which is the canonical YO structure, in which the short-wave block becomes ultralocal and the coupling to the long wave disappears entirely. What is lost in this limit is precisely what made the YON case hard, that is, the derivative term $\alpha\partial_x$ in the $(1,2)$ entry and the off-diagonal terms $\mp 2\ii \alpha \hS$. Solving $Q_0 \delta H = \text{(flow)}$ on the graded ansatz returns
\begin{equation}
    H_{\text{YO}} = \int \left[2L|S|^2 + \frac{2}{\beta}|S_x|^2 \right]\, dx.
\end{equation}
As apparent from this, the reduction does not work directly for the Hamiltonian. The Hamiltonian $H_*$ of Eq. \eqref{H*} features $\alpha$ in the denominators, up to $\alpha^{-3}$ through the term $\beta^2L/8\alpha^3$, so its YO reduction diverges. Therefore, the pair $(\PP_*,H_*)$ defining the Hamiltonian structure has no finite limit for YO, even though $\PP_*$ and the flow it generates do. The reason is clear from the form of $Q_0$: since its long-wave corner is $\partial_x$ and its long-short coupling is gone, $\delta \int L\,dx=(0,0,1)$ lies in its kernel. As $\alpha$ approaches 0, the Hamiltonian runs off in the direction of the Casimir.

The point symmetries change at the same place: $X_4$ is lost, and the scaling $X_5$ of the YO grading and the shift of $L$ with a time-dependent phase, $X_6$, appear instead, the latter being Hamiltonian with respect to $Q_0$: $Q_0 \delta \int (xL+2t|S|^2)\,dx=(-\ii\beta tS, \ii \beta t S^*,1)$.

For the second structure, the limit is again finite and nonzero, with $\frac14 \partial_x^3$ in the long-wave corner and $-\frac{\ii\beta}{2}\partial_x^2$ in the short-wave block. Most interestingly, for $\alpha=0$ the nonlocal tail of $\PP_2$ collapses to rank one: out of the six nonzero entries of $c$ only the $(1,1)$ entry survives the limit, leaving a tail of the form $-\frac{3\beta^2}{4} u_1\partial_x^{-1} u_1$, with $u_1=(S,-S^*,0)$, a single leg on the phase symmetry. This is exactly the shape of the tail of the second Hamiltonian structure of the NLS equation. The rank-three tail of Section \ref{SecTail} is therefore a genuinely two-parameter phenomenon, the legs of the flow and the translation being switched on by $\alpha$.

Finally, the conserved tower does not degenerate: the fifteen nontrivial densities produced by the recursion \eqref{DensitiesDiagonalisation} through order six span a space of gradients of rank six both for generic $\alpha$ and at $\alpha=0$, so that the six independent ones remain independent in the limit. This is worth stating because a natural-looking basis does collapse: the low-weight densities obey $\rho_1\rightarrow -|S|^2$ and $\rho_2\rightarrow -\beta|S|^2$ at $\alpha=0$ and become proportional there. That is a defect of the weight-by-weight basis for the densities and not of the tower itself; the Lax recursion produces a basis that does survive the limit.

\section{Conclusions}\label{SecConclusions}
To close out the paper, let us give a brief tally of the results presented and an outlook of the questions they leave open.

In this work we have studied the Hamiltonian and bi-Hamiltonian structure of the Yajima--Oikawa--Newell system, which is integrable for every choice of its two parameters. In Section \ref{SecNoGo}, we noted that, for $\alpha\neq0$, a constant shift of the long wave reduces the YON system to the Newell system, for which two Hamiltonian operators, both nonlocal, were given in \cite{Liu1994}, and we used the grading induced by the scaling symmetry and the $U(1)$ invariance of the system to reduce the search for local Hamiltonian operators to finitely many candidates at each weight, none of which is a Poisson operator for $\alpha\neq0$. In Section \ref{SecNonlocal}, we showed that the nonlocality of the first of those operators is tied to the choice of variables: a nonlocal gauge transformation on the potential of the long wave turns it into a local, non-degenerate operator $\PP_*$ for every $\alpha\neq0$ and $\beta$, which is the Lie--Poisson structure of the $\mathfrak{sl}(2)$ current algebra, the two signs of $\alpha$ corresponding to its two real forms. The gauge parameter is not a choice, but is fixed to $\gamma=-3\alpha$ by the Jacobi identity, and it differs from the gauge of Newell, $\gamma=\alpha$, which linearises the short-wave equation. The Hamiltonian formulation allowed us to make explicit the Noether correspondence between the Lie point symmetries of the system and its conserved quantities, the scaling symmetry included, and we also obtained the conserved densities of the system in closed form by asymptotic diagonalisation of the Lax pair. In Section \ref{SecSecond}, we constructed, in explicit weakly nonlocal form, a recursion operator advancing the hierarchy by two steps and, from it, a second Hamiltonian structure $\PP_2=\RR\PP_*$, lying above the pair of \cite{Liu1994} in the hierarchy, whose nonlocal tail is generated precisely by the three autonomous Lie point symmetries of the system, and we proved that $\PP_*$ and $\PP_2$ form a Hamiltonian pair for every $\alpha\neq0$ and $\beta$. Finally, in Section \ref{SecLenard}, we studied the Lenard--Magri chain of the pair. Since the recursion operator admits no square root and $\PP_*$ has no Casimirs, the chain splits into two interleaved families, which have to be seeded by the two simplest conserved quantities, $\int L\,dx$ and $\int\rho_1\,dx$. The flows of these seeds, the phase rotation and the translation, are symmetries of every member of the hierarchy, and that is enough to make every flow local and every Hamiltonian local and in involution, to all orders. For $\beta=0$, that is, for the Newell system, the flow is bi-Hamiltonian with respect to this pair, while for $\beta\neq0$ the YON system itself is bi-Hamiltonian up to a Galilean change of frame.

Some of these features are worth highlighting beyond the YON system. The double step of the recursion is inherited from the Yajima--Oikawa limit, but the deformation by $\alpha$ removes the Casimir that anchors the chain there: at $\alpha=0$, $\int L\,dx$ is a Casimir of the first structure, whereas for $\alpha\neq0$ it is not. The combination of a structure without Casimirs, a hierarchy without gaps and a recursion operator without square root makes the Lenard--Magri scheme a propagation mechanism rather than a generating one, the role of the Casimir being played by the symmetries generated by the seeds. Let us also note that the composition $\PP_2=\RR\PP_*$ fails on the constant gradient $\delta\int L\,dx$, a point that is easy to overlook and that affects the Lenard relations of every Hamiltonian with a component along $\int L\,dx$, which flows are Hamiltonian with respect to both structures, and the nonlocal form of the first structure in the original variables.

Several questions remain open. Whether the two families exhaust all the conserved densities of the system remains open. The reduction $\alpha=\beta=0$, a free Schr\"odinger equation driving a slaved long wave, admits no local Hamiltonian operator of weight one or two, and whether it admits a Hamiltonian structure at all is not known. It would also be natural to relate the pair to the Lax-operator brackets of the constrained KP and modified KP hierarchies \cite{Oevel1993}, and in particular to determine the image of $\PP_2$ under the Miura map of \cite{Liu1994}, which takes $\PP_*$ to the second Hamiltonian structure of the YO system; to understand the $\mathfrak{sl}(2)$ current algebra of the first structure in terms of the $3\times3$ Lax pair \eqref{LaxPair}; and to extend the construction to multicomponent versions of the YON system. Finally, the bi-Hamiltonian structure may prove useful in the analysis of the solutions of the system, in particular in connection with its inverse scattering problem, to which the standard techniques do not apply directly, the matrix $\Sigma$ in the Lax pair \eqref{LaxPair} being singular \cite{CasoHuertaThesis}.

\section*{Acknowledgments}
The author is grateful to Matteo Sommacal and Sara Lombardo for insightful discussions and for their help in proofreading the manuscript. This work has been carried out under the auspices of the Italian GNFM (Gruppo Nazionale Fisica Matematica), INdAM (Istituto Nazionale di Alta Matematica), which is gratefully acknowledged.

\bibliographystyle{acmunsrt}
\bibliography{biblio}

\end{document}